\documentclass[twoside,fleqn]{article}
\input epsf
\usepackage{latexsym}
\newcommand{\tilsub}[1] {\raisebox{-0.4ex}{$\stackrel{\textstyle 
\bf #1}{\scriptstyle \sim}$}}
\newcommand{\AmS}{{\protect\the\textfont2
  A\kern-.1667em\lower.5ex\hbox{M}\kern-.125emS}}

\def\beq{\begin{equation}} \def\eeq{\end{equation}}

\begin{document}
\title{Screening in (2+1)D pure gauge theory at high temperatures
       \thanks {Supported by Deutsche Forschungsgemeinschaft 
                under grants Pe-340/3-2 and Pe-340/6-1.}}
\author{E. Laermann, C. Legeland and B. Petersson\\
        Fakult\"at f\"ur Physik, Universit\"at Bielefeld\\
        Postfach 100131, D-33501 Bielefeld, Germany}

\maketitle

\begin{abstract}
We compute heavy quark potentials in pure gauge $SU(3)$ at high
temperatures in $2+1$ dimensions and confront them with expectations
emerging from perturbative calculations.
\end{abstract}

\section{INTRODUCTION}

The physical properties of the quark-gluon plasma still
present a largely unsettled problem insofar as
quantitative
information about screening lengths and quasi-particle
excitations is not really available yet.
Lattice
calculations provide interesting results, nevertheless, 
at very high temperatures
one would like to make contact with perturbation theory.

However, the domain of applicability of
perturbative calculations is not known. 
The main difficulty originates from
the infrared divergencies of the finite temperature
theory. 
In the chromoelectric sector they are supposed to be regulated
by the Debye mass, which appears in the resummation of the
polarization diagrams. If these diagrams dominate, one would expect
that the resummed perturbation theory gives a good description
of the potential between heavy quarks.
       
In order to investigate these questions of principle, we turn
to a simpler, but very similar theory, namely SU(3) gauge
theory in 2+1 dimensions. This theory still has confinement
at low temperatures, and a phase transition to a deconfined
high temperature phase. In early investigations, d'Hoker
both studied perturbation theory \cite{hok1} and performed some
Monte Carlo calculations on the analogous SU(2) model \cite{hok2}.
Other investigations of the SU(2) and SU(3) model in 
three dimensions have been devoted to the string tension
at zero temperature \cite{xxx} and the properties of the phase 
transition \cite{dam}. In the present contribution we describe the
first results of our high statistics investigation of the
SU(3) gauge theory in the high temperature phase. 
We test the relations expected from a naive
resummation of perturbation theory. In contrast to
the conclusions drawn by d'Hoker we
will show below that the description based on naively resummed
perturbation theory is not in agreement with the numerical data. 

\section{THE MODEL}

The lattice version of the three dimensional pure SU(3)
gauge theory, which we use, is given by the conventional
Wilson action.
The coupling constant $g^2$ in three dimensions has
non-trivial dimension so that, in the scaling region, 
all physical quantities with dimension mass 
are expected to have a constant ratio to $g^2$.
To determine the screening mass as a function of temperature in
the high temperature phase we study two quantities, namely
the cyclic Wilson loop and the correlation between Polyakov loops.
The cyclic Wilson loop $W(R)$ is the colour-averaged trace of
a closed rectangular loop of link
matrices along the contour shown in Fig.1 
where $R = \,\mid\vec{x} - \vec{y} \mid$ ist the spatial 
distance between the two Polyakov-loop like fractions of it.
From this quantity and the Polyakov loop $L(R)$ we define the potentials
\begin{equation}
V_{cwl}(R)=-\frac{<\, W (R) \,> - <\,L\,>^2}{<\,L\,>^2}\, T
\end{equation}
and
\begin{equation}
V_{plc}(R)=-\frac{<\,L(0) L^{\dag}(R) \,> - <\,L\,>^2}{<\,L\,>^2}\, T
\end{equation}

\begin{figure}[t]
   \begin{center}
      \leavevmode
      \epsffile{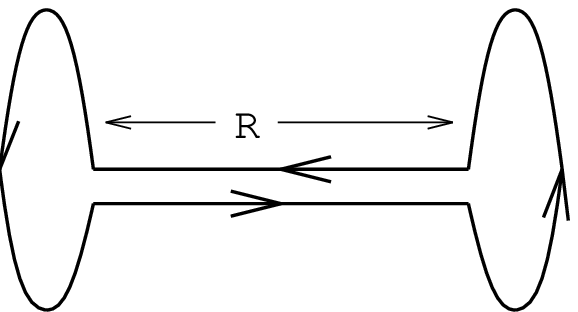}
      \label{fig:cwl}
      \caption{The cyclic Wilson loop}	              
    \end{center}
\end{figure}

In perturbation theory, one can choose a gauge, where the spacelike
links of the cyclic Wilson loops are set to one. Therefore, the
lowest order contribution from perturbation theory 
comes from a single gluon exchange
between the timelike sides of the rectangle. 
In contrast, for the correlation
of the Polyakov loops, the lowest order is the exchange of two
gluons, because the Polyakov loops are separately gauge singlets.
Higher order polarization graphs are strongly infrared divergent,
which can be taken into account by the selfconsistent introduction
of a screening mass. In leading order this simple recipe leads 
to the following expressions for the potentials, in $SU(N)$,
\beq
 V_{cwl}(r) = -\frac{N^2-1}{2N} \frac{1}{2\pi} K_0(m r)   
 \label{for:Vcwl}
\eeq
and 
\beq
 V_{plc}(r) = - (V_{cwl}(r))^2/16 
 \label{for:Vplc}
\eeq
where the screening mass $m$ is given by \cite{hok1}
\beq
 m^2 =\frac{g^2NT}{4\pi} \ln \frac{T}{g^2},
 \label{for:DHoker_scal_mu}
\eeq
and $K_0$ is the modified Bessel function. Note that we distinguish
between lattice and continuum quantities, $R=r/a$ and $\mu=ma$.

\section{CRITICAL TEMPERATURE}

The critical temperature was determined on lattices with time
extent $N_{\tau}=2, 4, 6$. A finite size scaling analysis 
from volumes $16^2, 32^2, 48^2, 64^2$ and $192^2$ gives,
with $\beta = 6/a g^2$,
\begin{eqnarray}
\beta_c=  8.163(8)   &           & N_{\tau}=2 \\
\beta_c= 14.705(58)  & {\rm for} & N_{\tau}=4 \\
\beta_c= 21.508(572) &           & N_{\tau}=6 
\end{eqnarray}
Thus, we find that the critical temperature scales
quite well, $T_c / g^2 = \beta_c / 6N_{\tau} \simeq const$.
By comparing with the string tension in the zero
temperature three dimensional theory, we find
\begin{equation}
\sqrt{\sigma}/T_c= 1.070(2) \;\; {\rm for} \;\; N_{\tau}=4
\end{equation}
This number is in astonishingly good agreement with the prediction 
from string theory \cite{olesen},
$\sqrt{\sigma}/T_c = \sqrt{\pi/3} = 1.023 $.

\section{SCREENING MASS}

The location of the phase transition 
defines a temperature scale, which we use, under
the assumption of scaling, to determine the potentials at
various values of $T/T_c=1.5,2,3,6,12,24$. 
For this we use
lattices with $N_{\tau}=4$ and $N_{\sigma}=32, 48$ and $64$
where the larger spatial extent has been employed at the
largest values of $T/T_c$, such that $N_{\sigma} \mu > 5 $.
As for the determination of the critical temperature, 
we performed typically six overrelaxation and one
pseudoheatbath step per sweep. 
For each temperature we have accumulated
between 50000 and 200000 sweeps. This data base resulted in
small statistical errors of the correlation functions
up to distances of 
$rT \tilsub{<} 4$. We first
tried to fit the data with eqs.(\ref{for:Vcwl}),(\ref{for:Vplc}), starting 
at $r_{min}=1/T$, ($R_{min}=N_{\tau}$).
We found that these fits did not properly represent the
shape of the correlation functions. The fit did not improve when
we employed the
finite lattice expression corresponding to the $K_0$-function.
Therefore we appplied a more general Ansatz
\beq
 V_{cwl} (R) = -\frac {const}{(\mu R)^{\gamma}} e^{-\mu R}
 \label{for:Vcwlexp}
\eeq
\begin{figure}[htbp]
   \begin{center}
      \leavevmode
      \epsffile{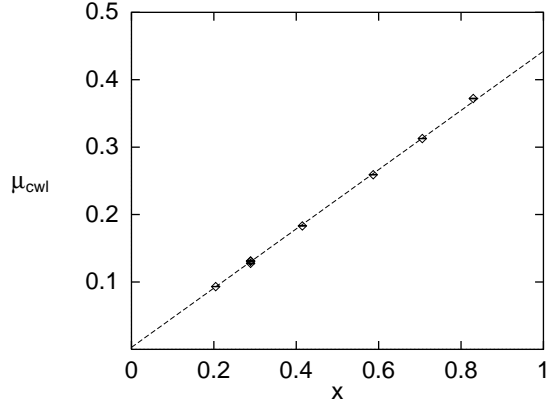}
      \caption{The screening mass (in lattice units)
               from cyclic Wilson loops, $\mu_{cwl}$, 
               as a function of temperature, $x = 1/{\rm sqrt}(T/T_c)$.}
      \label{fig:mu_cwl}
    \end{center}
\end{figure}
When we let both parameters $\gamma$ and $\mu$ vary freely, we obtain
values of $\gamma$ very near to 0.25 while
the asymptotic expansion of $K_0$ leads to $\gamma=0.5$.
Therefore we decided to employ eq.(\ref{for:Vcwlexp}) keeping 
$\gamma=0.25$ fixed
to extract the values for $\mu$. In Fig.\ref{fig:mu_cwl} we 
show $\mu$ plotted 
versus $x=1/\sqrt{T/T_c}$. The straight line denotes the 
fit result
\beq
 \mu = 0.0029(3) + 0.4395(6) x
 \label{for:scal_mu}
\eeq
For $N_{\tau}$ fixed, this corresponds to a dependence of the
physical mass $m$ proportional to $\sqrt{T}$, as given by
eq.(\ref{for:DHoker_scal_mu}). We see, however, no sign of the 
logarithmic correction which is only expected to dominate 
for $\mu \ll 1$. 

The correlation of the Polyakov loops, $V_{plc}$, is again well
described by a formula like eq.(\ref{for:Vcwlexp}), with
$\gamma=0.5$ and $\mu$ replaced by $2\mu$ in the exponent.
Also here eq.(\ref{for:Vplc}) is definitely less good. 

\begin{figure}[htbp]
   \begin{center}
      \leavevmode
      \epsffile{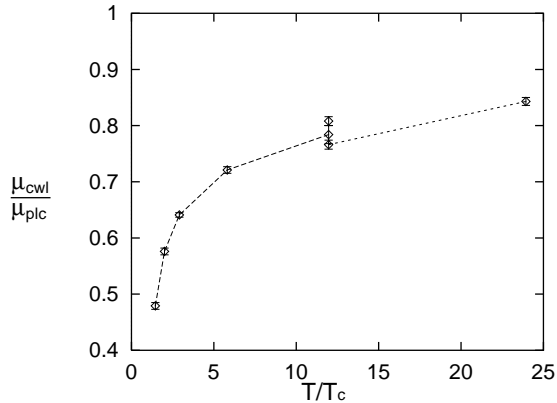}
      \caption{The ratio between the screening masses 
      $\mu_{plc}/\mu_{cwl}$
      as a function of
      temperature. The three results at $T = 12 T_c$ are obtained
      from three different lattices with $N_{\sigma} = 32,48$ and $64$.}
      \label{fig:mu_T}
    \end{center}
\end{figure}

It is still very interesting to see whether the basic
assumption regarding the exchange of one versus two 
single particles with fixed mass holds, even 
if the simple, resummed perturbative formulae,
eqs.(\ref{for:Vcwl}),(\ref{for:Vplc}), do not fit the data.
Therefore, we 
check the screening masses from the cyclic Wilson loop
and from the Polyakov loop correlation for equality.
In Fig.\ref{fig:mu_T} we plot
the ratio between those two masses which should be equal to one
if factorization holds. 
What we find, however, is a ratio of about $1/2$ at $T$ values 
slightly bigger than $T_c$, and only a slow approach to one
at very high temperatures, $\mu_{plc}/\mu_{cwl} \simeq 0.85$
at a temperature as high as $T = 24 T_c$.

\begin{figure}[htbp]
   \begin{center}
      \leavevmode
      \epsffile{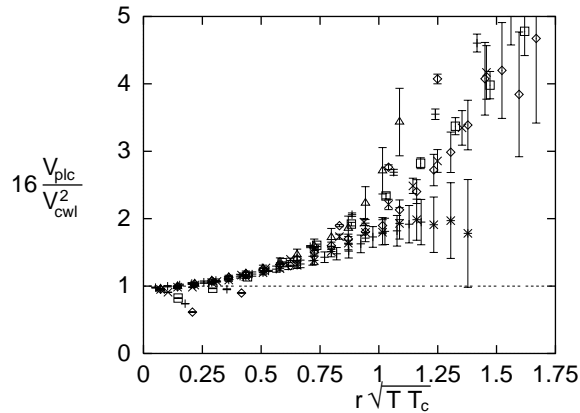}
      \caption{The ratio $16V_{plc}/V_{cwl}^2$ as a function
      of scaled physical distance $r {\rm sqrt}(T T_c)$. Different
      symbols denote different temperatures.}
      \label{fig:VVrt}
    \end{center}
\end{figure}
Another test of the factorization hypothesis involves a
direct comparison of the potentials, i.e. a check of
eq.(\ref{for:Vplc}).
In Fig.\ref{fig:VVrt} we plot the ratio $16V_{plc}/V_{cwl}^2$ 
as a function of
$r \sqrt{T T_c}$. At short distances and high temperatures this
ratio goes to one, as given by eq.(\ref{for:Vplc}), 
but at larger distances
this behaviour is no longer seen, although the various ratios
seem to scale in the variable $r \sqrt{T}$. 
We therefore conclude that a more
complicated mechanism, involving bound states between
the gluons is at work. We are further investigating this
possibility.


\begin{thebibliography}{99}

\bibitem{hok1}
E. D'Hoker, Nucl. Phys. B180 (1981) 341.

\bibitem{hok2}
E. D'Hoker, Nucl. Phys. B200 (1982) 517.

\bibitem{xxx}
M. L\"utgemeier, this conference.

\bibitem{dam}
J. Christensen, G. Thorleifsson, P.H. Damgaard and J.F. Wheater, 
Nucl. Phys. B374 (1992) 225.

\bibitem{olesen}
P. Olesen, Phys. Lett. B160 (1985) 408.

\end{thebibliography}
\end{document}